# A corrector design using active vase mirrors that allows a fixed telescope to access a large region of the sky.

Gilberto Moretto and Ermanno F. Borra

*Center for Optics, Photonics and Lasers - COPL, Physics Department, Laval University, Québec, QUE, G1K7P4, Canada. G. Moretto is now with the Steward Observatory, The University of Arizona, Tucson, AZ 85721 - USA.*

## Abstract

We investigate a family of two-mirror correctors to compensate the aberrations of a parabolic mirror observing at a large angle from the zenith. We constrain our designs to optical elements that can be built with presently available technology. The secondary and tertiary mirrors are warped using Zernike polynomials which we know can be generated with active vase mirrors. The performances of these corrector designs are useable for imagery.

## 1. Introduction

It is well known that the equilibrium configuration of the surface of a spinning liquid takes the shape of paraboloid. By using a reflecting liquid, mercury for example, it is thus possible to make a parabolic mirror. Liquid mirrors are one to two orders of magnitude less expensive than optical quality glass mirrors.

The feasibility and progress of liquid mirrors has been documented by Borra *et al.*[1] who built and tested a diffraction limited 1.5-metre liquid mirror (LM) followed by a 2.5-metre (Borra *et al.*[2]). Hickson *et al.*[3] have built a 2.7-metre f/2 liquid mirror telescope (LMT), obtaining images with FWHM < 2 Arcsec. Liquid mirrors are also interesting for non-astronomical applications. Atmospheric science physicists have used liquid mirrors as receivers for LIDAR applications. For example, a LIDAR system using a 2.65-metre LMT has been built and operated for some time (Sica *et al.*[4]). At NASA Lyndon B. Johnson Center, Potter *et al.*[5] have built a 3-meter LMT that is now in operation at Cloudcroft Observatory, New Mexico, and is used for detecting space debris.

One of the often cited limitations of liquid mirror telescopes pertains to the small region of sky which they can observe. Because the aberrations of a parabola increase rapidly with field angle, classical corrector designs cannot yield subarcsecond images for angles significantly greater than one degree. For example, the UBC/Laval 2.7-metre LMT (Hickson *et al.*[3]) has a classical lens corrector, designed by C. Morbey, that uses 5 lenses and yields subarcsecond images over a 0.8 Deg field. To access larger fields, innovative corrector designs must be explored. A landmark design, using a fixed primary mirror, was given by Richardson & Morbey[6] who devised a three-mirror system that permits a 10-metre, f/5 LMT to be operated at 7.5 Deg from the zenith. Borra[7] has explored analytically how far off axis one can use a liquid mirror telescope and has shown that, in principle, aberrations can be corrected for zenith distances as large as 45 Deg. We have recently begun exploring two-mirror designs to correct the aberrations of a parabolic primary mirror observing at a large zenith angle. Two-mirror correctors have previously been proposed by a number of designers, Paul[8], Lagrula[9] and Baker[10] with a number of variations, resulting finally in the well known Paul-Baker corrector (Angel *et al.*[11]). A first exploration (Borra, Moretto & Wang[12], hereafter referred to as paper I) investigated a two-mirror corrector used with a parabolic primary mirror observing at large zenith angles. The design, dubbed BMW corrector, uses complex surfaces that give subarcsecond images as far as 22.5 Deg from the optical axis. However, there then arises the practical problem of making the required complex anamorphic aspheric surfaces with existing technology .

In this article, as a further step toward a design feasible with existing technology, we further investigate the BMW design assuming mirrors that can be made with the *Active Vase Mirror* technology pioneered by Lemaître[13] [14].

## 2. Active Vase Mirrors

As a first step towards a corrector feasible with existing technology, we have investigated the aspherization of an active vase mirror for wavefront corrections up to third and fifth-order aberrations. We tested a stainless steel AISI 420 prototype mirror having a 16 cm aperture, a controlled pressure load, and two series of 12-punctual radial force application positions distributed symmetrically in two concentric rings around the mirror. Details relative to the theory and elasticity design of *Active Vase Mirrors* were presented by Lemaître and Wang[15]. Optical tests of a prototype active vase mirror were presented by Moretto *et al.*[16].

The equation for surface deflection for each zone $w$ is given by a 4th order differential equation (cf. Timoshenko & Woinowsky-Krieger[17]).

$$\nabla^2 \nabla^2 w\ (r, \theta) = q/D\ , \tag{1}$$

with $D = Et^3/[12(1-\nu^2)] = constant$, where $\nabla^2 w$ is the Laplacian of the flexure; $q$ is the uniform load; E and $\nu$ are Young's modulus and Poisson's ratio respectively; $t_1$ and $t_2$ and $D = D_1$ for $0 < r < a$, $D = D_2$ for $a < r < b$ are respectively the thicknesses and associated rigidities for the inner ($r = a$) and outer ($r = b$) zones of the active mirror.

The classes of the deformation terms feasible with vase mirrors can be obtained from the solutions for Eq. (1) when the value of the load $q$ is zero or constant. Figure 1 summarizes the low-order solutions of Eq. 1 where the shaded terms are the solutions that are feasible with the vase mirrors[18]. These terms, demonstrated experimentally by Moretto *et al.*[16] are those which we use to warp the vase mirrors in the design optimizations considered in this article.

### 3. BMW Correctors warped with Zernike polynomials

It is a happy coincidence that elasticity theory gives solutions expandable in series of Zernike polynomials that are easy to make, for the optical aberrations of an optical system can also be expressed with Zernike polynomials. In what follows we shall explore BMW correctors having secondary and tertiary mirrors warped with Zernike polynomials. We shall use the notation of Born & Wolf[19] and Malacara[20]. We use an initially spherical surface, on which we add aspherical terms by means of Zernike polynomials. This is done because spherical mirrors are less expensive than elliptical or parabolic mirrors. The choice of Zernike polynomials depends on the symmetry of the system. As shown in Figs 3 and 5 we have bilateral symmetry with respect to the Y-Z meridional plane which means that the system is symmetrical in the X-component but not on the Y-Z meridional plane; therefore the optical axes of the correctors and the optical axis of the primary mirror (i.e. optical axis of the system) are coplanar in the Y-Z meridional plane.

Using only the available terms presented above and assuming preservation of the bilateral symmetry, we have considered secondary and tertiary mirrors warped with the Zernike terms given in Tab. 1. All of the polynomials in Tab. 1 were demonstrated experimentally by Moretto *et al.*[16]. Defocus $[C_6(2\rho^2 - 1)]$ is not considered in the optimization since it can be compensated by changing the curvature of the vase mirrors.

### 4. Corrector Designs for a 4-m Diameter Fixed Primary Mirror Observing Off-Axis

We will consider zenith pointing telescopes having 4-m parabolic primary mirrors observing off-axis. As discussed in Paper I, the BMW design is very versatile; in this article we select two designs among those considered by Moretto[18]. To ensure adequate sampling (2.355 pixels per FWHM) with the pixel sizes of typical CCD detectors, we impose a spatial image sampling of 0.6 Arcsec/pixel for a pixel size of 25 microns in the first design and 30 microns for the second one, imposing correctors that yield telescope scales of respectively 20 and 24 Arcsec/mm and angular fields of view (FOV) of 12 and 18 Arcmin respectively. This corresponds to observations of point sources displaced from the reference positions by ±0.10 or ±0.15 Deg in the orthogonal direction. Figure (2) shows a lay-out of the (FOV) with nine positions $f_{1...9}$, where XAN and YAN are the angular distances in X and Y coordinates respectively. On this plane we have the focal plane or surface image ($S_i$) defined by the axis $X_{Local}$ and $Y_{Local}$ with the $X_{Local} - axis$ displaced in the Y(YAN)-direction. We take advantage of the bi-lateral symmetry setup defining field points including only the positive or negative X-component. With this symmetry ray tracing[21] is only done for half of the mirror pupil.

As explained in Paper I, a BMW corrector could, in principle, track in real time by warping and moving the mirrors; however, given the present state of the technology, it is more realistic to consider fixed mirrors and an electronic tracking system that uses a CCD driftscanning in a time delayed integration (TDI) mode. Hickson *et al.*[3] show an image taken with a liquid mirror telescope and a CCD tracking with the TDI technique. With the TDI technique pixels move on a straight line so that the distortion introduced by the corrector degrades the PSF. Therefore, it is important that the distortion introduced by the corrector be kept small. We also constrained our mirrors to maximum diameters of the order of 1 meter, since we feel confident that active vase mirrors can be built at these diameters.

**A. A Corrector Design for a Primary Mirror Observing 5.0 Deg from the Zenith Having a FOV of 12 Arcmin.**

We first consider a corrector having two vase mirrors for a 4-m f/4.5 parabolic primary mirror observing 5.0 Deg from the zenith that gives a field of view of ±0.1 Deg. As explained by Borra, Moretto & Wang[12], a fixed corrector set to observe at an angle $\theta$ from the zenith can observe anywhere inside $2\theta$. The relatively large f ratio would not increase much the cost of the telescope and observatory since the telescope is simply a fixed tower and the observatory a silo with sliding roof. To ensure adequate sampling, as discussed above, we impose a telescope scale of 20 Arcsec/mm resulting in a telescope effective focal length (EFL) of 10313 mm. Note that unlike the designs considered in paper I, we utilize the entire vase mirrors and not only sections of mirrors. The Zernike polynomials are centered on the optical axes of the vase mirrors. The geometrical parameters of the system are presented in Fig. 3, where the dimensions are given in millimetres. The values of Zernike coefficients for the secondary and tertiary vase mirrors are presented in Tab. 2.

Figure 4 shows the PSFs at Y = 4.90, 5.00, 5.10 Deg from the zenith displaced from each position by ±0.10 Deg in the X-direction. The correctors have been optimized to give good images simultaneously for the 9 spots. The *50 %*, *80 %* and *100 %* encircled energy diameters (EED) and the RMS spot diameter (RMS-SD), are shown in Tab. 3. To restrict the corrector distortion to an acceptable maximum value of 1.10 Arcsec P-V as shown in Tab.. 4, all the while preserving the image resolution, e.g. RMS values in the 0.604 - 0.74 Arcsec range with a field of view of 12 Arcmin, some optical concessions are required: a small amount of curvature in the image plane ($R_{Img} = 21843.69$) as well as secondary and tertiary active mirrors having maximum diameters of 1.05 metres. These requirements restrict our design to a zenith distance of 5 Deg and a field of 12 Arcmin. In the next section we shall use additional refractive elements placed in front of the focal plane to improve performance and to better control distortion. Note that a system tracking in real time would not be restricted by distortion and would have better performance.

**B. A Corrector for a Primary Mirror Observing 7.50 Deg from the Zenith with a Field Of View of 18 Arcmin.**

We now explore a corrector design for a *4.0 m f/5.25* Liquid Mirror observing at 7.5 Deg from the zenith having a FOV of 18 Arcmin. Again we impose a spatial image sampling of 0.6 Arcsec per 25 *μm* pixel size, which results in a telescope scale of 24 *Arcsec/mm*. To obtain good images we must work with a slower primary mirror. To better control the distortion we introduce three additional lenses, the largest one having spherical surfaces and the other two aspherical surfaces, as shown in Figs. 5, 6 and 7. The optimizations were done for the wavelength band 600 to 800 nm. The geometrical parameters, in millimetres, are given in Fig. 5 and the values of Zernike coefficients for the secondary and the tertiary vase mirrors are presented in Tab. 5. The geometrical parameters for the three-lens set, surfaces $S_4$ to $S_9$ are presented in Tab. 6 and 7. We use one spherical and two aspherical lenses, where the sag Z of the aspherical surface parallel to the surface's optical axis is given by

$$Z = \frac{c\rho^2}{1 + \{1 - (1-K)c^2\rho^2\}^{1/2}} + A\rho^4 + B\rho^6 + C\rho^8 + D\rho^{10} , \qquad (2)$$

where c = 1/R is the curvature at the pole of the surface, R is the radius of curvature, K is the conic coefficient, A,B,C, and D are the 4th, 6th, 8th and 10th order deformation coefficients respectively and $\rho^2 = x^2 + y^2$.

The PSFs computed at Y = 7.35, 7.50, 7.65 Deg from the zenith and displaced from each position by ±0.15 Deg in the X-direction give the 9 spots, optimized for the wavelength band 600 to 800 nm, are shown in Fig. 8 (a) for 600 nm, (b) for 700 nm, (c) for 800 nm. The *50 %, 80 %* and *100 %* encircled energy diameters (EED) and the RMS spot diameter (RMS-SD) are shown in Tab. 8. Figure 9 shows encircled energy plots. The corrector distortion control is presented in Tab. 9. As we can see, the addition of one spherical and two aspherical lenses enables us to increase the observation angle, measured from the zenith, from 5.0 to 7.5 Deg and the FOV from 12 to 18 Arcmin, with RMS values in the 0.555 − 0.782 Arcsec range and with a control of the corrector distortion to a maximum value of 0.145 Arcsec P-V.

Using the same initial conditions presented in this section we explore a design which could cover a wider wavelenght band. The optimizations were done in the wavelength band 400 - 700 nm. The geometrical positions of each element and the distortion control are very similar to those for the last design. To the 3-lens group we have used three aspherical lenses with conic constant ($K_n = 0$) for the lens' surfaces (n=4, ... , 9). The radial energy distribution of the geometrical spot diagram for each field in terms of the *50 %, 80 %* and *100 % encircled energy diameters (EED)* and the *RMS spot diameter (RMS-SD)* are shown in Tab. (12).

## 5. Discussion

This is the latest in a series of papers that study corrector designs for fixed telescopes such as liquid mirror telescopes. While the first few articles were preliminary explorations, culminating with the BMW design, the latest papers, including the present one, specifically address the feasibility of such correctors. In this paper we show that practical correctors having reasonable performances can be made by mechanically warping spherical mirrors. The designs considered in this paper only incorporate warping modes that we have demonstrated experimentally with a small prototype active vase mirror (Moretto *et al.*[16]).

The performance of these correctors is measured by how far off-axis one can use a fixed primary mirror and how large is the field of view with subarcsecond PSFs. At this stage, driftscanning is the only practical tracking method: As a consequence, distortion must be minimized. This, in turn, imposes additional constraints that limit the performance of the corrector. Adding refractive elements improves distortion but renders the design more complex. With real time tracking, we could tolerate distortion, and could work with simpler better performing designs. Unfortunately real time tracking involves real time warping of the mirrors as well as real time changes in the geometry of the correctors. This certainly can be implemented but it is premature to envision it, given the present state of the technology. Using only one type of glass (BK7-Schott) for the lens-group, the achromatic aberration lies within a tolerable limit. If less achromatic aberration is required a new lens group with different glasses could be explored in order to decrease the achromatism.

These designs were explored for regions of up to 7.5 Deg off axis all the while limiting the secondary and the tertiary mirror diameters to less than about 1 meter and using only Zernike terms which are feasible using the current active vase mirror technology. Using larger mirrors or using off-axis sections of anamorphic mirrors (Paper I) yields better images and increases the area accessible to the fixed primary mirror. Prima facie, an accessible field of 15 Deg seems small when compared to the nearly 180 Deg that are in principle accessible to a conventional telescope. Close scrutiny however shows a more favorable situation. In practice, conventional telescopes seldom observe at zenith distances greater than 45 Deg since atmospheric absorption increases and seeing deteriorates rapidly beyond that limit. Benn & Martin[22] have compiled statistics about the use of the Herschel telescope, finding that 94% of the observations are made within 50 Deg of the zenith. Taking this into consideration and based on the fact that we have corrector designs that can extend the accessible field for a fixed telescope to more than 45 Deg, we can assert that fixed telescopes, such as liquid mirrors telescopes, can be competitive with tiltables ones. This can readily be seen by noting that, even limiting the design to ± 7.5 Deg, at a terrestrial latitude of 30 Deg, a strip of sky 15-degree wide and 24 hours long centered at the zenith contains 4,600 square Deg of sky. This corresponds to 11% of the entire sky (both hemispheres) and 18% of the sky observable with a conventional telescope, at same site, limited within 45 Deg of the zenith. All along it must be kept in mind that LMTs, even equipped with BMW correctors, are considerably less expensive than conventional telescopes.

To assess the applicability of warping, one must consider the maximum amplitudes of the deflections and the maximum stresses allowed by the materials. In other words, we need to know if the active mirror design and material can produce the deflexions required by our designs. We have considered secondary and tertiary mirrors warped with the terms: $Z_5$, $Z_{10}$, $Z_{11}$, $Z_{13}$, $Z_{14}$ and $Z_{21}$, where $Z_n = C_n F(\rho, \theta)$, presented in Tab. 1. These terms are fitted on an initial spherical shape ($C_1 = 0$), as was discussed in Section 3. Based on these Zernike terms and the curvatures of the secondary and the tertiary mirrors for the two designs presented, we obtain maximum amplitudes of deflection $w_{max}$ at the edges of these mirrors. The values for these maximal amplitudes of deflection $w_{max}$, in millimetres, for each Zernike term $Z_i$ are presented in Tabs. 10 and 11. The astigmatism term $[Z_5 = C_5 \, \rho^2 \cos 2\theta]$ for the tertiary mirror in the $7.50 \pm 0.15 Deg$ design has the largest maximum deformation amplitude value, i.e. +0.91584 mm. As was done by Lemaître & Wang[15], we take as a first approach the astigmatism coefficient to be

$$C_5 \approx (1+\nu)\sigma_{max}/Et_2, \tag{3}$$

where $\sigma_{max}$ is the maximal stress; $E = 2.05 \ 10^4 \ daN \ mm^2$ is Young's modulus; $\nu = 0.305$ is the Poisson ratio corresponding to Fe-Cr13, which is the material used in the active mirror; and $t_2 = 12 \ mm$ is the axial thickness of the outer ring of the active mirror. With the coefficient $C_5 = 3.8144 \ 10^{-06} \ mm^{-1}$ for the tertiary mirror in the $7.50 \pm 0.15$ Deg design, the maximal stress is $\sigma_{max} = 0.719036 \ daN \ mm^2$. The maximum tolerable stress for Fe-Cr13 is 50 $daN \ mm^2$, as discussed in Lemaître & Wang[15]. This indicates that the maximum amplitude of deformation value, $w_{max} = +0.91584 \ mm$ lies within the elastic limit of the active mirror. Furthermore, as was demonstrated in Moretto[18], the deformation limit corresponding to the *Astm3* term, using the 20-mm active vase mirror, was 2.4 mm. This confirms that it is possible to produce the corrections proposed in this series of correctors.

## 6. Conclusion

We have investigated a family of two-mirror correctors to compensate the aberrations of a parabolic mirror observing at a large angle from the zenith. Our main goal is to provide fixed telescopes, such as liquid mirror telescopes, access to as much of the visible sky as possible as well as large enough fields of view (FOV) to be useful for imagery. We have constrained our designs to optical elements that can be built with presently available technology. Therefore, the secondary and tertiary mirrors were warped using Zernike polynomials which we know can be generated with active vase mirrors. The performances of these corrector designs are useable for imagery. We show that the surfaces generated by the vase mirrors are indeed suitable for making the required corrections for fixed parabolic mirrors observing off-axis. The result is a practical design that uses existing technology.

We show that distortion can be minimized with the help of an additional three-lens ele-ment. Because CCDs driftscan along straight lines, distortion must be minimized, imposing additional constraints on our designs. Note also that for stationary telescopes, not located on the equator, stars move along curved paths at varying speeds. Maximum spatial resolutions are obtained when the stars move in straight lines and at constant speed along each pixel columns within the FOV and with the same value of speed for each column. As pointed out by Richardson[23], it is possible to introduce asymmetrical distortion to compensate the star trail curvature and the differential sidereal rate to improve the resolution of the images obtained with a driftscanning CCD. Given the present state of the technology, it is premature to consider real time tracking; driftscanning with a CCD detector is far more realistic, Hickson[3] et al and Zaritsky[24] et al.

In terms of giant telescopes, i.e. larger than 10-meters, it is mechanically simpler to keep the primary mirror immobile, pointing directly overhead and use reflective correctors to access the desirable zenith distances. Also, it should be noted that a classical tiltable telescope can only observe one field at a time. A fixed primary mirror telescope, on the other hand, could simultaneously access many widely separated fields with several correctors.

One must realize that the designs presented here are but an additional step toward even better performing correctors. In this article we simply present solutions which can be done with existing technology. We are confident that better practical designs can be found.

## ACKNOWLEDGMENTS

This research has been supported by the Natural Sciences and Engineering Research Council of Canada. G. Moretto was also supported by CAPES and FAPESP (SP) - Brazil. We are grateful to Prof. E.H.Richardson and Prof. G.R. Lemaître for useful discussions.

# REFERENCES

1. E.F. Borra, R.Content, and L.Girard, S. Szapiel, L.M.Tremblay, and E. Boily, "Liquid Mirrors: Optical Shop Tests and Contributions to the Technology", The Astrophysical Journal **393**, 829-847 (1992).

2. E.F. Borra, R.Content, and L.Girard, "Optical Shop Tests of a f/1.2 2.5-meter Diameter Liquid Mirror", The Astrophysical Journal **418**, 943-946, (1993).

3. P. Hickson, E.F. Borra, R. Cabanac, R. Content, B. K. Gibson, and G.A.H. Walker, "UBC/Laval 2.7 meter Liquid Mirror Telescope", The Astrophysical Journal, **436**: L201 - L204, (1994).

4. R. J. Sica, S. Sargoytchev, P.S. Argall, E.F. Borra, L. Girard, C. T. Sparrow, and S. Flatt, "Lidar measurements taken with a large aperture liquid mirror. 1. Rayleigh-scater system", App. Opt., Vol. 34, **30**, (1995).

5. B. Iannotta, (Journalist), "Spinning mages from mercury mirrors", New Scientist, **147**, 38-41, No 1986, July 1995.

6. E.H. Richardson & C.L. Morbey, "Fifteen Degree Correcting Optics for a 10-Meter Liquid Mirror Telescope", Instrumentation for Ground-Based Optical Astronomy Present and Future, p.720, ed. L. B. Robinson (New York: Springer), (1987).

7. E.F. Borra, "On the Correction of the Aberrations of a Liquid Mirror Telescope Observing at Large Zenith Angles", Astronomy and Astrophysics, Astronomy and Astrophysics, **278** 665-668 (1993).

8. M. Paul, Rev. Opt. **14**, 169 (1935).

9. N.J. Rumsey, *Optical Instruments and Techniques*, Oriel, Newcastle, (1970).

10. J.G. Baker, "On Improving the Effectiveness of Large Telescopes", IEEE Transactions of Aerospace and Electronic Systems **AES-5**, 261 (1969).

11. J.R.P. Angel, N.J. Woolf. and H.W. Epps, "Good images with very fast paraboloidal primaries: an optical solution and some applications". S.P.I.E. Conference Proceedings - International Conference on Advanced Technology Telescopes **332**, 134 (1982).

12. E.F. Borra, G.Moretto, M. Wang, "An optical corrector design that allows a fixed telescope to access a large region of the sky", Astronomy and Astrophysics Suplementary Series, **109**, 563-570 (1995).

13. G. R. Lemaître, "Active Optics and Elastic Relaxation Methods", Current Trends in Optics, International Commission for Optics, Taylor & Francis Publ., London, 135 (1981).

14. G. R. Lemaître, *Various Aspects of Active Optics*, Proc. SPIE Conf. on Telescopes and Active Systems, Orlando, FA, 328 (1989).

15. G. R. Lemaître and M. Wang, "Active Vase Mirrors Warped by Zernike Polynomials for Correcting Off-Axis Aberrations of Fixed Primary Mirrors. – Part 1 - Theory and Elasticity Design", Astronomy and Astrophysics Suplementary Series, **114**, 373-378 (1995).

16. G. Moretto, G.R. Lemaître, T. Bactivelane, M. Wang, M. Ferrari, S. Mazzanti, B. Di Biagio, and E. F. Borra, "Zernike Polynomials for Correcting Off-Axis Aberrations of Fixed Primary Mirrors. – Part 2 - Optical Testing and Performance Evaluation", Astronomy and Astrophysics Suplementary Series, **114**, 379-386 (1995).

17. Timoshenko, S., Woinowsky-Krieger, S., *Theory of Plates and Shells*, (McGraw-Hill International, 1959).

18. G. Moretto, Ph.D. Thesis:*Optical Corrector Design for Fixed Primary Mirrors Observing Off-Axis.*, Laval University, Québec - Que, Canada - (1996).

19. M. Born & E.Wolf, *Principles of Optics*, 3rd ed.(Pergamon Press, New York,1964).

20. D. Malacara, *Optical Shop Testing*, (John Wiley and Sons. - N. York, 1978).

21. We use the software CodeV to the optical optimizations. CodeV is a trademark of Optical Research Associates, California, USA, (1995).

22. C.R. Benn, R. Martin, QJRAS **28**, 481, (1987).

23. E.H. Richardson, *"Corrector Lens Design for UBC 5-metre Liquid Mirror Telescope" - Private Communication*, (1995)

24. D. Zaritsky, S. A. Shectman, G. Bredthauer, "The Great Circle Camera: A New Drift-Scanning Instrument", Publications of the Astronomical Society of the Pacific, **108**: 104-109, 1996.

FIGURES

Fig. 1.   Solutions of Eq. (1) when the value of the load $q$ is zero or constant. The shaded terms are the classes of the deformation terms that are feasible with vase mirrors.

Fig. 2.   The lay-out shows the angular field of view (FOV) with nine positions $f_{1...9}$ defined on the plane X(XAN) and Y(YAN). On this plane we have the focal plane or surface image ($S_i$) defined by the axes $X_{Local}$ and $Y_{Local}$ with $X_{Local} - axis$ displaced in the Y(YAN)-direction. The quantities $\Delta(X_n - X_m)$ and $\Delta(Y_n - Y_m)$ are the differences, in millimetres, of the chief-ray's X and Y coordinates for the positions $f_n$ and $f_m$.

Fig. 3.   This diagram shows a schematic design for a 4-m diameter f/4.5 mirror observing at $\theta = 5.0$ Deg from the zenith and with a FOV of 12 Arcmin. $\phi$ represents the diameter in millimetres of each mirror and $R_n$ is the radius of curvature for the n-th surface. The system is symmetrical in the X-Z plane. The dimensions are given in millimetres. The human figure at the right bottom of the telescope gives a handy reference scale.

Fig. 4.   Spot diagrams for point sources observed at 4.90, 5.00 and 5.10 Deg from the zenith and displaced by ± 0.10 Deg in the orthogonal direction. The parameters of the corrector have been optimized to give good images simultaneously for the 9 spots.

Fig. 5.   This diagram shows a schematic design for a 4-m diameter f/5.25 mirror with the correctors observing at $\theta = 7.50$ Deg from the zenith and with a FOV of 18 Arcmin. $\phi$ represents the diameter for each mirror and $R_n$ is the radius of curvature for the n-th surface. The dimensions are given in millimetres. Note that the system is symmetrical in X-Z plane. The details for the three lens set are shown in Figs. 6 and 7. The human figure at the right bottom of the telescope gives a handy reference scale.

Fig. 6.   This figure illustrates details of the 3-lens group and the active secondary and tertiary mirrors for the 4-m diameter f/5.25 mirror observing at $\theta = 7.50$ Deg from the zenith and with a FOV of 18 Arcmin.

Fig. 7.   The above shows the details of the 3-lens group for the system, where THI is the distance between subsequent surfaces measured along the optical axis of the lens, YDE is the surface decenter along the Y-direction of the lens and ADE is the tilt of the lens' Y-axis in the Y-Z plane. Each decenter defines a subsequent surface displaced and/or rotated from the precedent. The dimensions are given in millimetres and presented in Tab. 6.

Fig. 8.   Spot diagrams for point sources observed at 7.35, 7.50 and 7.65 Deg from the zenith and displaced by ± 0.15 Deg in the orthogonal direction, at wavelengths (a) 600 $nm$, (b) 700 $nm$, and (c) 800 $nm$. The parameters of the corrector have been optimized for the wavelength band 800 - 600 nm.

Fig. 9.   The Encircled Energy Diameter (EED) for point sources observed at 7.35, 7.50 and 7.65 Deg from the zenith and displaced by ± 0.15 Deg in the orthogonal direction. Because of the bilateral symmetry, the curves for +0.15 Deg are the same as those for -0.15 Deg.

TABLES

Table 1. Standard Zernike polynomials used in the optical optimizations.

| Zernike Polynomial | Name |
|---|---|
| $C_5$ $\rho^2 \cos 2\theta$ | *Astigmatism with axis at $\pm \pi/4$* |
| $C_{10}$ $(3\rho^3 - 2\rho) \sin \theta$ | *Primary Coma in x-direction* |
| $C_{11}$ $\rho^3 \sin 3\theta$ | *Triangular Coma5* |
| $C_{12}$ $\rho^4 \cos 4\theta$ | *Square Astigmatism7* |
| $C_{13}$ $(4\rho^4 - 3\rho^2) \cos 2\theta$ | *Secondary Astigmatism* |
| $C_{14}$ $(6\rho^4 - 6\rho^2 + 1)$ | *Primary Spherical* |
| $C_{21}$ $(5\rho^5 - 4\rho^3) \sin 3\theta$ | *Triangular Coma7* |

Table 2. The values of Zernike coefficients for the secondary and tertiary vase mirrors for the 5.00 ±0.10 Deg design.

| Term | Secondary Mirror $M_2$ | Tertiary Mirror $M_3$ |
|---|---|---|
| $C_5$ | $-5.1164 \ 10^{-07}$ | $-4.7739 \ 10^{-08}$ |
| $C_{10}$ | $+7.5126 \ 10^{-10}$ | $-5.2866 \ 10^{-11}$ |
| $C_{11}$ | $+3.3932 \ 10^{-11}$ | $+4.0868 \ 10^{-13}$ |
| $C_{13}$ | $-1.5286 \ 10^{-14}$ | $-2.1068 \ 10^{-14}$ |
| $C_{14}$ | $-1.2791 \ 10^{-13}$ | $-1.7411 \ 10^{-14}$ |
| $C_{21}$ | $-4.8947 \ 10^{-18}$ | $+2.7071 \ 10^{-18}$ |

Table 3. Parameters that characterize the correction at 5.0 Deg from the zenith: a single two-mirror corrector corrects at ±0.1 Deg and at the wavelength of 632.8 nm.

| Field(x,y) | RMS-SD[a] | EED[b] [Arcsec] | | |
|---|---|---|---|---|
| [Deg] | [Arcsec] | 50 % | 80 % | 100 % |
| (0.00, 4.90) | 0.711 | 0.389 | 0.710 | 2.490 |
| (0.00, 5.00) | 0.619 | 0.322 | 0.553 | 2.211 |
| (0.00, 5.10) | 0.627 | 0.307 | 0.624 | 1.676 |
| (± 0.10, 4.90) | 0.743 | 0.421 | 0.816 | 2.545 |
| (± 0.10, 5.00) | 0.604 | 0.357 | 0.638 | 2.643 |
| (± 0.10, 5.10) | 0.658 | 0.428 | 0.701 | 1.935 |

[a]RMS spot diameter.

[b]Encircled Energy Diameter.

Table 4. Corrector Distortion at 632.8 nm for each section of FOV that characterizes the corrections for the selected zenith angles: $\theta = 5.00 \pm 0.10$. $X_n$ and $Y_n$ are the field's X and Y-distances, taking as a reference the lay-out presented in Fig. 2.

| X-Section | Distortion in Arcseconds | Y-Section | Distortion in Arcseconds |
|---|---|---|---|
| $\Delta(X3-X1)$ | +0.000 | $\Delta(Y6-Y3)$ | -0.022 |
| $\Delta(X6-X5)$ | +0.571 | $\Delta(Y5-Y2)$ | -0.036 |
| $\Delta(X6-X4)$ | +1.113 | $\Delta(Y4-Y1)$ | -0.050 |

Table 5. The values of Zernike coefficients for the secondary and tertiary vase mirrors for 7.50 ±0.15 Deg design.

| Term | Secondary $M_2$ | Tertiary $M_3$ |
|---|---|---|
| $C_5$ | $+3.5290 \; 10^{-06}$ | $+3.8144 \; 10^{-06}$ |
| $C_{10}$ | $+1.2209 \; 10^{-09}$ | $+4.7994 \; 10^{-10}$ |
| $C_{11}$ | $-3.3318 \; 10^{-11}$ | $+8.9266 \; 10^{-11}$ |
| $C_{13}$ | $+8.9039 \; 10^{-14}$ | $+3.3165 \; 10^{-14}$ |
| $C_{14}$ | $-6.0444 \; 10^{-13}$ | $-4.3788 \; 10^{-13}$ |
| $C_{21}$ | $+1.3052 \; 10^{-17}$ | $+2.2727 \; 10^{-17}$ |

Table 6. Geometrical parameters for the three-lens set, surfaces $S_4$ to $S_9$. The dimensions are given in millimetres. The value of circular apertures (APE) measures the illuminated portion of the surface.

| $S_n$ | Radius [mm] | THI [mm] | YDE [mm] | ADE [Deg.] | APE [mm] | Glass Schott |
|---|---|---|---|---|---|---|
| $S_4$ | -294.48 | -64.92 | 2895.82 | 10.00 | 250.00 | BK7 |
| $S_5$ | -234.37 | -92.55 | — | — | — | — |
| $S_6$ | -1439.28 | -40.00 | -70.00 | 24.00 | 150.00 | BK7 |
| $S_7$ | -922.64 | -88.88 | — | — | — | — |
| $S_8$ | -1341.68 | -30.00 | +57.62 | -11.12 | 90.00 | BK7 |
| $S_9$ | +329.84 | -50.00 | — | — | — | — |
| $S_i$ | INFINITY | 0.00 | 0.00 | -3.59 | — | — |

Table 7. The values of asperics terms for the second ($S_6$&$S_7$) and the third ($S_8$&$S_9$) lenses. K is the conic coefficient and A,B,C, and D are the 4th, 6th, 8th, 10th order deformation coefficients, respectively.

| $S_n$ | K | A | B | C | D |
|---|---|---|---|---|---|
| $S_6$ | 0.00 | $+0.39\ 10^{-07}$ | $+0.88\ 10^{-12}$ | $-0.98\ 10^{-16}$ | $+0.41\ 10^{-20}$ |
| $S_7$ | 0.00 | $+0.46\ 10^{-07}$ | $+0.11\ 10^{-11}$ | $-0.14\ 10^{-15}$ | $+0.62\ 10^{-20}$ |
| $S_8$ | 61.20 | 0.00 | 0.00 | 0.00 | 0.00 |
| $S_9$ | -2.93 | $-0.43\ 10^{-09}$ | $-0.21\ 10^{-11}$ | $+0.70\ 10^{-15}$ | $-0.71\ 10^{-20}$ |

Table 8. Parameters that characterize the correction at 7.50 Deg from the zenith: a single two-mirror corrector corrects at ±0.15 Deg.

| $Field(x,y)$ [Deg] | RMS-SD[a] [Arcsec] | | | EED[b] [Arcsec] | | |
|---|---|---|---|---|---|---|
| | 600 nm | 700 nm | 800 nm | 50 % | 80 % | 100 % |
| (0.00, 7.35) | 0.670 | 0.286 | 0.461 | 0.384 | 0.814 | 1.224 |
| (0.00, 7.50) | 0.629 | 0.390 | 0.657 | 0.454 | 0.781 | 1.671 |
| (0.00, 7.65) | 0.890 | 0.401 | 0.423 | 0.409 | 0.999 | 1.453 |
| (± 0.15, 7.35) | 0.957 | 0.539 | 0.466 | 0.502 | 0.941 | 1.870 |
| (± 0.15, 7.50) | 0.695 | 0.440 | 0.647 | 0.497 | 0.834 | 1.783 |
| (± 0.15, 7.65) | 0.913 | 0.531 | 0.557 | 0.556 | 0.991 | 1.861 |

[a]RMS spot diameter.

[b]Encircled Energy Diameter for the wavelength band 600-800 nm.

Table 9. Corrector Distortion at two wavelength for each section of FOV that characterizes the correction at selected zenith angles: $\theta = 7.50 \pm 0.15$ Deg. $X_n$ and $Y_n$ are the fields' X and Y-distances, taking as a reference the lay-out presented in Fig. 2.

| *FOV* | *Distortion in Arcseconds* | | *FOV* | *Distortion in Arcseconds* | |
|---|---|---|---|---|---|
| *X-Section* | *600 nm* | *800 nm* | *Y-Section* | *600 nm* | *800 nm* |
| $\Delta(X3-X1)$ | 0.000 | 0.000 | $\Delta(Y6-Y3)$ | 0.129 | 0.145 |
| $\Delta(X6-X5)$ | 0.113 | 0.128 | $\Delta(Y5-Y2)$ | 0.123 | 0.133 |
| $\Delta(X4-X5)$ | 0.129 | 0.127 | $\Delta(Y4-Y1)$ | 0.126 | 0.128 |

Table 10. The maximum amplitudes of deflection $w_{max}$ for the $7.50 \pm 0.15 Deg$ design, in millimetres, for each Zernike term $Z_i$, where $\rho = 465$ mm for the secondary and $\rho = 490$ mm for the tertiary active mirrors.

| *Zernike Terms $Z_i$* | *Secondary Mirror* $w_{max}$ [mm] | *Tertiary Mirror* $w_{max}$ [mm] |
|---|---|---|
| $Z_5 = C_5 \, \rho^2 \cos 2\theta$ | +0.76306 | +0.91584 |
| $Z_{10} = C_{10} \, (3\rho^3 - 2\rho) \sin \theta$ | +0.36826 | +0.16939 |
| $Z_{11} = C_{11} \, \rho^3 \sin 3\theta$ | -0.00335 | +0.01050 |
| $Z_{13} = C_{13} \, (4\rho^4 - 3\rho^2) \cos 2\theta$ | +0.016651 | +0.00765 |
| $Z_{14} = C_{14} \, (6\rho^4 - 6\rho^2 + 1)$ | -0.16955 | -0.15145 |
| $Z_{21} = C_{21} \, (5\rho^5 - 4\rho^3) \sin 3\theta$ | +0.00142 | +0.00321 |

Table 11. The maximum amplitudes of deflection $w_{max}$ for the 5.00±0.10$Deg$ design, in millimetres, for each Zernike term $Z_i$, where $\rho = 550$ mm for the secondary and $\rho = 550$ mm for the tertiary active mirrors.

| Zernike Terms $Z_i$ | Secondary Mirror $w_{max}$ [mm] | Tertiary Mirror $w_{max}$ [mm] |
| --- | --- | --- |
| $Z_5 = C_5\ \rho^2 \cos 2\theta$ | -0.15477 | -0.01444 |
| $Z_{10} = C_{10}\ (3\rho^3 - 2\rho) \sin \theta$ | +0.37497 | -0.02638 |
| $Z_{11} = C_{11}\ \rho^3 \sin 3\theta$ | -0.00335 | +0.00414 |
| $Z_{13} = C_{13}\ (4\rho^4 - 3\rho^2) \cos 2\theta$ | +0.00565 | -0.00771 |
| $Z_{14} = C_{14}\ (6\rho^4 - 6\rho^2 + 1)$ | -0.07023 | -0.00956 |
| $Z_{21} = C_{21}\ (5\rho^5 - 4\rho^3) \sin 3\theta$ | -0.00123 | +0.00068 |

Table 12. Radial energy distributions for the spot diagrams which characterize the corrections at the selected zenith angles: $\theta = 7.50\ \pm 0.15$ Deg. The optimization was done for the wavelength band 400 - 700 $nm$.

| Field(x,y) [Deg] | RMS-SD [Arcsec] | EED [Arcsec] | | |
| --- | --- | --- | --- | --- |
| | | 50 % | 80 % | 100 % |
| (0.00, 7.35) | 0.693 | 0.381 | 0.802 | 1.776 |
| (0.00, 7.50) | 0.634 | 0.335 | 0.690 | 1.760 |
| (0.00, 7.65) | 0.806 | 0.513 | 0.958 | 1.897 |
| (± 0.15, 7.35) | 1.003 | 0.607 | 1.164 | 2.752 |
| (± 0.15, 7.50) | 0.765 | 0.446 | 0.784 | 2.213 |
| (± 0.15, 7.65) | 0.948 | 0.709 | 1.206 | 1.882 |